\documentclass[CEJP,PDF]{cej} 

\usepackage{graphicx}

\usepackage{layout}
\usepackage{amsmath}
\usepackage{textcomp}
\usepackage{hyperref}

\title{Spectral and kinetic properties of electroluminescence of ZnS:Cu powder in polymer structure}

\articletype{Research Article}

\author{E~Chimczak\inst{1}\email{eugeniusz.chimczak@put.poznan.pl},
        T~Dunaj\inst{1},
        M~Bertandt\inst{1},
	A~Wieczorek\inst{1},
	G~Neunert\inst{2},
	G~Chimczak\inst{3},
	M~Cie{\.z}\inst{4},
	M~{\L}ukasik\inst{4}}

\institute{
     \inst{1} Poznan University of Technology, Faculty of Technical Physics,\\
              ul. Nieszawska 13A, 60-965 Pozna{\'n}, Poland
     \inst{2} Agricultural University, Department of Physics,\\
              ul. Wojska Polskiego 38/42, 60-637 Pozna{\'n}, Poland
     \inst{3} Adam Mickiewicz University, Nonlinear Optics Division, Department of Physics,\\
	      ul.Umultowska 85, 61-614 Pozna{\'n}, Poland
     \inst{4} Institute of Electron Technology,\\
	      ul. Zab{\l}ocie 39, 30-701 Krak{\'o}w, Poland
          }

\abstract{Spectral and kinetic measurements of the light output have been made for AC electroluminescent structure. ZnS:Cu  is luminescence active layer in the structure. In kinetic measurements, excitation was by rectangular wave voltage pulse of 1~ms duration. During the excitation the structure emits blue-green light. The maximum of the spectrum lies at about 455~nm.}


\begin{document}
\maketitle


\section{Introduction}

Electroluminescence is the phenomenon being subject of interest to many researches. During several decades very many papers were devoted to the phenomenon. The light emission from silicon carbide crystals excited by an applied voltage was first reported by Lossev in 1923~\cite{lossev23}. In 1936 Destriau made the electroluminescent cell based on zinc sulphide~\cite{destriau36}. At the end of fifties years Thornton was started work on the electroluminescent devices with vacuum deposited semiconductor layer~\cite{thornton59}. In the seventies much attention has been paid to doubly insulated AC thin electroluminescent devices for flat panel display~\cite{inoguchi74,suyama82}. In 1990 many researches have focused on polymer light-emitting diodes~\cite{burroughes90}. They are rather concerned with spectral, electrical and chemical properties of the diodes. In the present paper we are concerned with spectral as well as kinetic properties of the structure investigated.

\section{Experimental}

\begin{figure}
\includegraphics[width=0.7\textwidth]{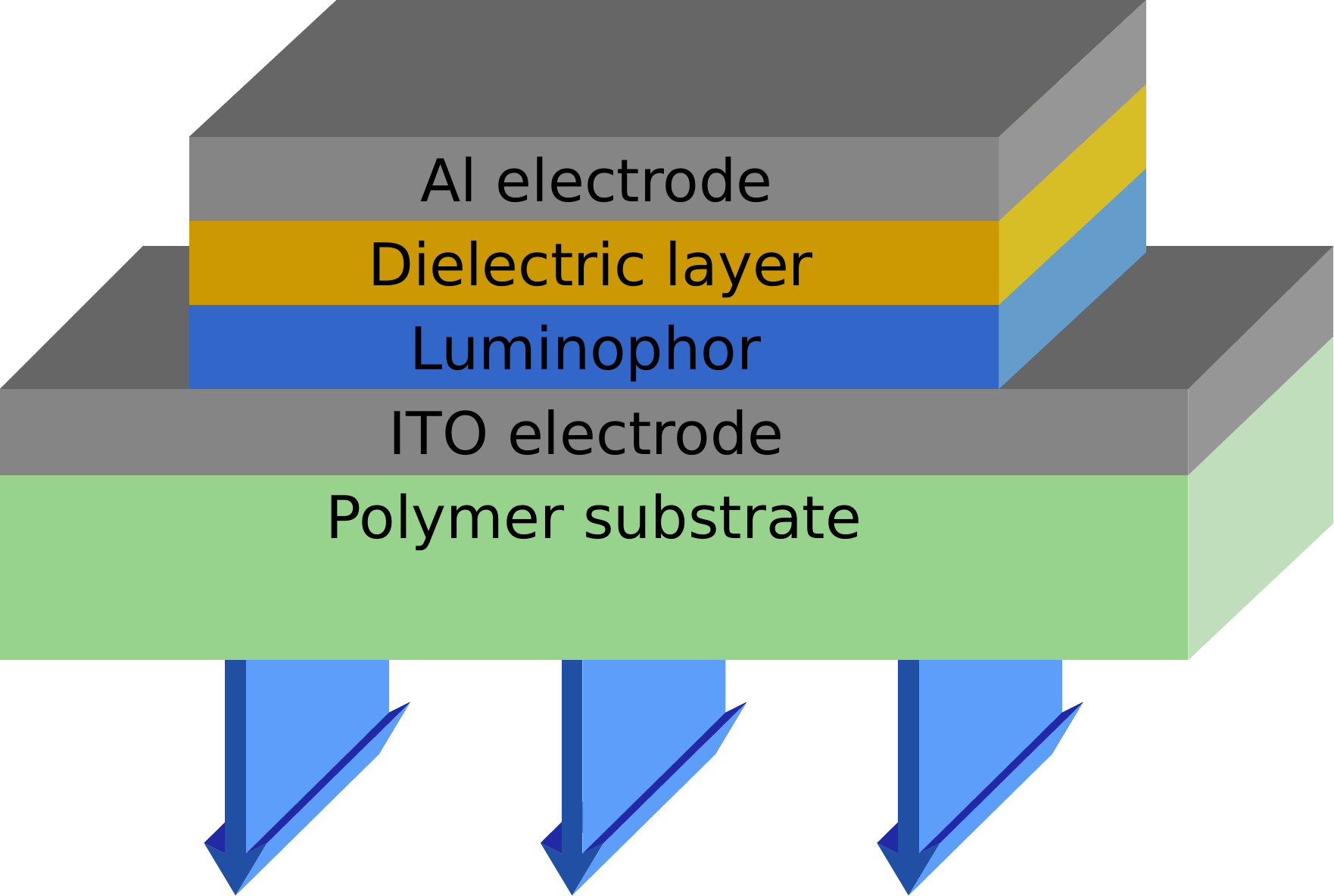}
\caption{Electroluminescent polymer structure used in the present study.\label{fig1}}
\end{figure}

Figure~\ref{fig1} shows the structure of the cell investigated. The structure consists of polymer substrate with deposited transparent electrode, luminophor in polymer matrix, dielectric layer (${\rm{BaTiO_{3}}}$ powder in polymer matrix) and Al electrode. ITO( Indium-Tin-Oxide) was used as transparent electrode. A zinc sulphide activated with copper at the concentration of $0.1$~wt.\% was used as the luminophor. Spectral measurements were made using a Centronic Q-4283BM photomultiplier connected to a Zeiss SPM2 grating monochromator. All spectra were corrected for photomultiplier sensitivity. Kinetic measurements were performed, exciting the samples by rectangular wave voltage pulse of 1~ms duration. The measurements were performed at room temperature.

\section{Results and discussion}

\begin{figure}
\includegraphics[width=0.7\textwidth]{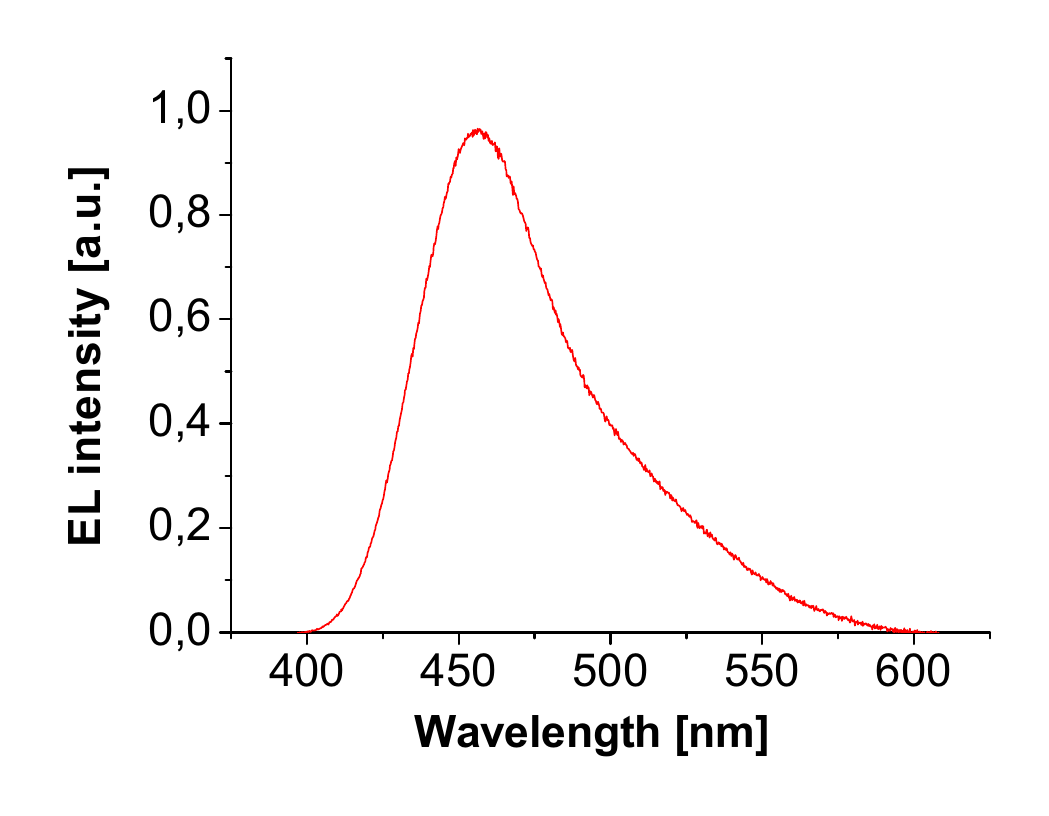}
\caption{Electroluminescence spectrum of the polymer structure.\label{fig2}}
\end{figure}

The electroluminescent structure investigated emits blue-green light. The spectrum of the electroluminescence is shown in figure~\ref{fig2}.
The electroluminescence was excited by AC voltage of 77.5~V at frequency of 2000~Hz. The spectrum  attains its maximum at about 455~nm.
The shape of the spectrum points that the spectrum consists of -- at least -- two elemental bands. Allieri {\em et al.}~\cite{allieri02} have been investigated the electroluminescent material (ZnS:Cu) 
embedded in a polymer matrix between thin film electrodes. These authors were mainly concerned with the spectral and electrical properties of their EL devices. They have been recorded the EL spectra at different voltages(from 40~V to 400~V) and frequencies (from 100~Hz to 2~kHz). They observed two bands at about 460~nm (blue region) and 510~nm (green region). The authors found that at low excitation frequency the green emission is more intense, while over 1~kHz the blue band predominates over the green one. In the light of Allieri {\em et al.}’ results’, our suggestion, that spectrum shown in figure~\ref{fig2} consists of two elemental bands, is probable. The position of the blue band maximum of our structure is in good accordance with the results obtained by Allieri {\em et al.}

\begin{figure}
\includegraphics[width=0.7\textwidth]{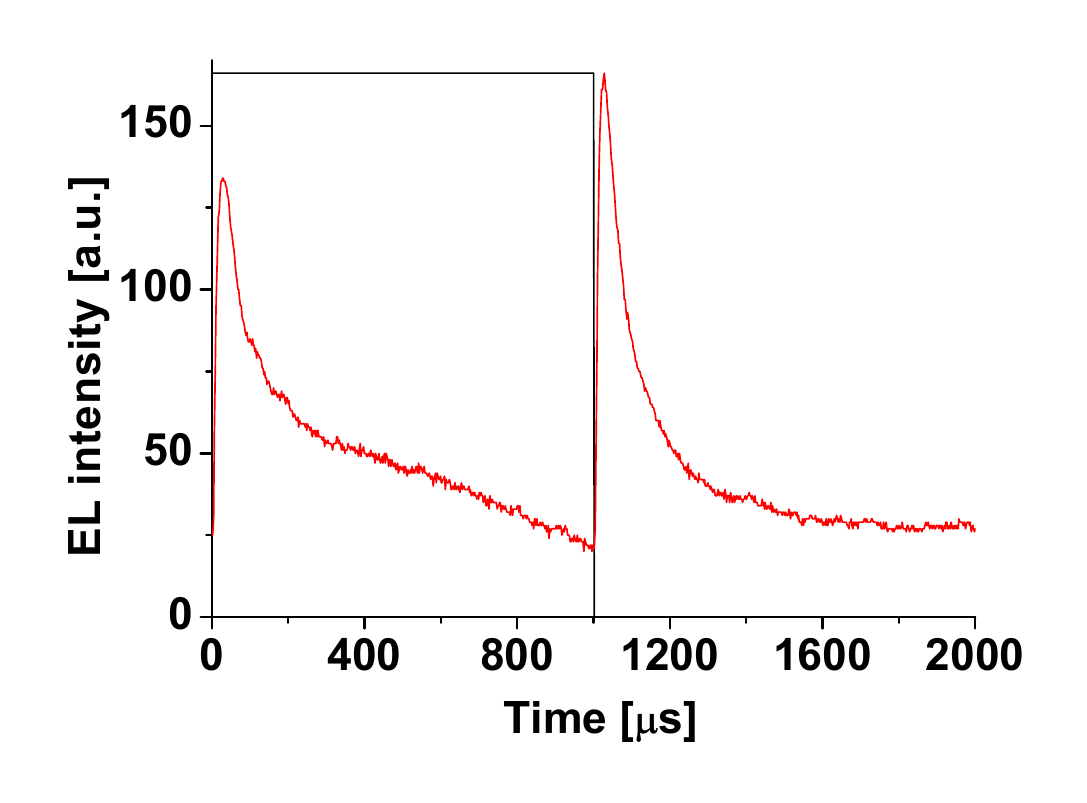}
\caption{Time dependence of the electroluminescence  of the polymer structure.\label{fig3}}
\end{figure}

\begin{figure}
\includegraphics[width=0.7\textwidth]{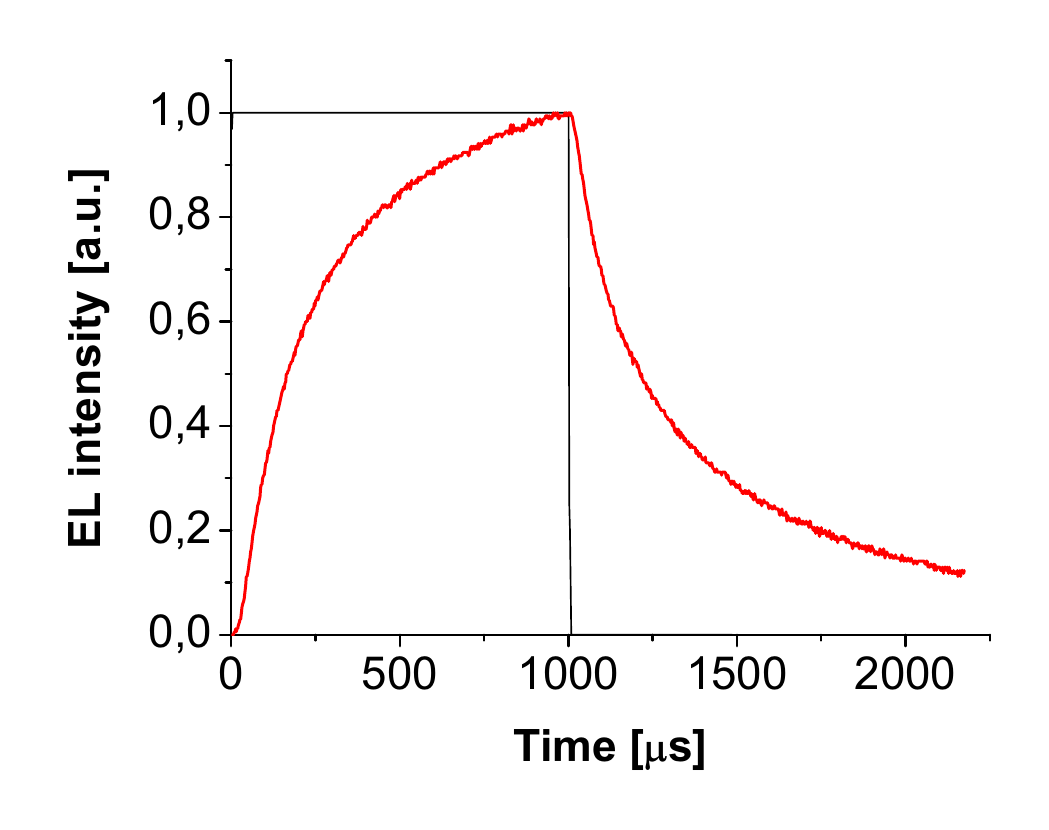}
\caption{Time dependence of the DC electroluminescent cell.\label{fig4}}
\end{figure}

In the luminescence investigations, kinetic measurements are very important because they provide much information concerning the luminescent centers and the mechanism of luminescence in general~\cite{dejene,chimczak88,chimczak88b,chimczak89,ravi04}. Especially important are the measurements of luminescence excited by a constant pulse, for example, voltage, during as well as after the end of the pulse~\cite{chimczak88,chimczak88b,chimczak89,ravi04}. Figure~\ref{fig3} shows the time dependence of luminescence of the structure investigated during and after the end of the  voltage pulse of  75~V, 1~ms.
As is seen in the figure, the electroluminescence quickly attains its maximum and then subsequently decays during as well as after the voltage pulse. It is well-known that, in the case of direct excitation of luminescent centers by rectangular pulse, the time dependence of the luminescence is described as

\begin{eqnarray}
\label{eq:exc_pul}
I_d&=&G\Big(1-e^{-\frac{t}{\tau}}\Big) \, ,
\end{eqnarray}
during  the exciting pulse and
\begin{eqnarray}
\label{eq:end_pul}
I_a&=&I_0 e^{-\frac{t}{\tau}} \, ,
\end{eqnarray}
after the end of the pulse, where $\tau$ is the lifetime of luminescent center and $G$ is the product of generation rate and luminescence yield. When there is energy transfer from another centre then the electroluminescence is described by 
\begin{eqnarray}
\label{eq:e3}
I_d&=&\frac{G}{\tau_2-\tau_1}\Big[\tau_2\Big(1-e^{-\frac{t}{\tau_2}}\Big)-\tau_1\Big(1-e^{-\frac{t}{\tau_1}}\Big)\Big] \, ,
\end{eqnarray}
and
\begin{eqnarray}
\label{eq:e4}
I_a&=&\frac{G}{\tau_2-\tau_1}\Big[\tau_2\Big(1-e^{-\frac{T}{\tau_2}}\Big) e^{-\frac{t}{\tau_2}}-\tau_1\Big(1-e^{-\frac{T}{\tau_1}}\Big) e^{-\frac{t}{\tau_1}}\Big] \, ,
\end{eqnarray}
where $\tau_1$ and $\tau_2$ are the lifetimes of the transferring and emitting centers and $T$ is exciting
pulse duration~\cite{chimczak88,chimczak88b}. Contrary to the direct excitation, the curve described by equation~(\ref{eq:e3}) has got a
point of inflection. Both the curves still increase during the exciting pulse. The above 
equations explain very well kinetic behaviour of many different luminescent cells. In figure~\ref{fig4}, for example,  is shown  time dependence of electroluminescence of ZnS:Mn,Cu thin film.
The beginning of the kinetic curve during the exciting pulse points that there is energy transfer to manganese centers from other centers (there is a point of inflection). So, in this case, we can assume that the constant voltage pulse is cause of the kinetic observed. Unfortunately, the above equations do not explain kinetics of the polymer structure investigated. From figure~\ref{fig3} one can see that the voltage pulse exciting our structure is not the direct cause of the electroluminescence time dependence.

\section{Conclusion}

Electroluminescence of polymer structure based on ZnS:Cu was presented. The structure was excited by rectangular wave voltage pulse of 1~ms duration. During the excitation  the blue-green light was observed. The spectrum of the electroluminescence attains its maximum at about 455~nm. The electroluminescence appears quickly after the beginning and the end of the voltage pulse and then subsequently decays. Mechanism of direct excitation  and energy transfer of luminescence excited by constant pulse can not be used in mathematical description of the electroluminescence.

\section*{Acknowledgment}

This work was supported by the Poznan University of Technology Grant No DS 62-176/12.

%


\end{document}